\documentclass[aps,rmp,reprint,amsmath,amssymb,graphicx,longbibliography]{revtex4-1}

\usepackage{chemformula} 
\usepackage{xfrac}  
\usepackage{float} 
\usepackage{physics}
\usepackage{dsfont}

\usepackage{graphicx}
\usepackage{dcolumn}
\usepackage{bm}
\usepackage{xcolor}
\usepackage{soul}

\begin{document}

\title{Decoherence of solid-state spin qubits: a computational perspective}
\author{Mykyta Onizhuk}
   \email{onizhuk@uchicago.edu}
   \affiliation{Pritzker School of Molecular Engineering, University of Chicago, Chicago, IL 60637, USA}
\author{Giulia Galli}%
   \email{gagalli@uchicago.edu}
   \affiliation{Pritzker School of Molecular Engineering, University of Chicago, Chicago, IL 60637, USA}
   \affiliation{Department of Chemistry, University of Chicago, Chicago, IL 60637, USA}
   \affiliation{Materials Science Division and Center for Molecular Engineering, Argonne National Laboratory, Lemont, IL 60439, USA}

\date{\today}

\begin{abstract}
    The usefulness of solid-state spins in quantum technologies depends on how long they can remain in a coherent superposition of quantum states. This Colloquium discusses how first-principles simulations can predict spin dynamics for different types of solid-state electron spins, helping design novel and improved platforms for quantum computing, networking, and sensing. 
    We begin by outlining the necessary concepts to understand the noise affecting generic quantum systems. We then delve into recent advances in predicting spin-phonon relaxation of spin-defect qubits. Next, we discuss cluster methods as a means of simulating quantum decoherence induced by spin-spin interactions, emphasizing the critical role of validation in ensuring the accuracy of these simulations. We highlight how validated cluster methods can be instrumental in interpreting recent experimental results and, more importantly, predicting the coherence properties of novel spin-based quantum platforms, guiding the development of next-generation quantum technologies.
\end{abstract}
\maketitle
\tableofcontents{}

\section{Introduction}
Isolated single electron spins in solids offer a transformative platform for quantum science and technology. For example, the realization of long-lived quantum states of spin qubits in a solid-state environment enables the implementation of high-fidelity gates for universal quantum computations~\cite{PRXQuantum.5.010102, Mills2022, Noiri2022, Xue2022, Madzik2022, Abobeih2022fault, Huang2024}, facilitating an efficient exploration of complex problems intractable by classical hardware~\cite{Google2019, Daley2022}. In addition, the inherent sensitivity to magnetic fields makes spin qubits ideal platforms for ultra-high resolution magnetometry \cite{hong2013, RevModPhys.92.015004, RevModPhys.96.025001}, while spin-charge coupling enables electric field sensing at the nanoscale \cite{Dolde2011, PhysRevApplied.16.024024}. Interfacing electron spins with photons allows for the creation of hybrid spin-photon systems as key components of quantum networks \cite{Stas2022, ruf2021, Pompili2021, Chen2019}, where nuclear spins act as memory units for long-term storage of quantum states~\cite{PhysRevX.6.021040, PhysRevX.9.031045}.

All aforementioned applications of single electron spins hinge on a key property - the time it takes for a qubit quantum state to deteriorate, referred to as \textit{coherence time}. In quantum networking, extended coherence times allow for the creation and preservation of entanglement between spins, enabling secure communication between distant nodes \cite{Hermans2022}; in quantum computation, they ensure the qubit maintains its quantum state to perform complex logical operations \cite{Yoneda2017}, and in sensing, long coherence times enable exquisitely high sensitivity in measurements of the environmental fluctuations \cite{Herbschleb2019}.

Accurate predictions of the coherence time of solid-state spin qubits are thus crucial for the development of next-generation qubit platforms and they may be obtained using ab initio theoretical and computational methods. Such approaches allow one to understand the interplay between spin dynamics, the interactions present in the host material, and possibly external forces, and eventually design systems with a tailored electronic structure and controllable decoherence mechanisms. 

In this Colloquium, we discuss recent developments in characterizing and understanding the dynamics of solid-state spin qubits using ab initio simulations. We begin with introducing spin-qubits in the solid state in Sec. \ref{sec:types}. Sec. \ref{sec:spin_relaxation} provides intuitive, simplified models to understand relaxation and dephasing mechanisms, respectively. In Sec. \ref{sec:theory}, we summarize methods to simulate these processes, with more in-depth discussion on spin-phonon relaxation in Sec. \ref{sec:phonons}, and on spin-spin decoherence in Sec. \ref{sec:spins}. Sec. \ref{sec:applications} highlights recent work on validating, interpreting, and predicting decoherence dynamics of solid-state spin qubits in semiconductors and insulators.

\section{Physical realization of solid-state spin qubits}\label{sec:types}
There are a plethora of ways to confine a single or a few spins into solid-state platforms. We can loosely characterize them into the following categories:
quantum dots, shallow donors, spin defects, and emerging platforms.

We include in the ``emerging platforms" category the spins of single electrons at the interface of condensed phases of noble gases \cite{PhysRevA.74.052338, PhysRevLett.105.040503, Chen2022}, confined single spins in carbon nanotubes \cite{Chen2023} and fullerenes \cite{Pinto2020}, electron spins in molecules \cite{Wasielewski2020, Zadrozny2015} and other platforms not covered by the list mentioned above. Many concepts discussed in this Colloquium are applicable to different spin qubit realizations; however, the applicability of first-principles simulations to investigate the electronic structure of spin systems greatly varies, as we discuss below.

For a comprehensive review of recent advancements in various types of qubits, including platforms other than spin qubits, we refer the reader to the recent publication by Cheng et al. \cite{Cheng2023}.

\subsection{Quantum dots}
Quantum dots are primary candidates for spin-based quantum computing applications \cite{PhysRevA.57.120}. To manufacture quantum dots in semiconductors, one uses a static electric field bias to spatially confine a single or few electronic spins \cite{RevModPhys.95.025003}. Such electrostatic engineering is achieved by restricting the electron density to interfaces between two semiconductors or quantum wells formed in heterostructures, and by controlling the electric field with metallic gates. A device based on spin qubit quantum dots is schematically shown in Figure \ref{fig:semiq}(a). The exemplary structure of the device is adapted from the recent work on a three-qubit array in silicon (ref. \cite{Takeda2021}).

Quantum dot-based technologies are operated at millikelvin temperatures to control thermal excitations. 
The first quantum dots in semiconductors were primarily investigated at the interface of \ch{GaAs}/\ch{AlGaAs}~\cite{Fujisawa2002}, but more recently the field has moved towards mostly silicon and silicon-germanium-based quantum dots. The readout is usually achieved via spin-to-charge conversion. There are various implementations of quantum dots-based spin qubits (single spin qubits, exchange qubits, singlet-triplet qubits, etc.), and an extensive review is reported by Burkard and coworkers \cite{RevModPhys.95.025003}.

An atomistic first-principles characterization of quantum dots of realistic sizes is not yet possible, as the square modulus of the electron wavefunction spans tens of nanometers in the material (Fig.~\ref{fig:semiq}a); therefore theoretical descriptions often adopt simplified semi-empirical approaches to model the electronic structure of quantum dots.

\subsection{Shallow donors}
Shallow donors in semiconductors and insulators are created, for example, by adding an atom of a group-V element, such as phosphorus or arsenic, to a group-IV crystal, e.g. silicon \cite{RevModPhys.85.961}. The donor atom replaces a host atom, and its extra electron is loosely bound to the donor nucleus in a hydrogen-like orbital. Arguably the most widely used system is \ch{^{31}P} donor in Si~\cite{Chatterjee2021}. The \ch{^{31}P}-dopants in silicon have an energy level of only 45 meV below the conduction band~\cite{PhysRev.140.A1246}; thus, all applications require cryogenic temperatures. In quantum technologies, both the nuclear spin-\sfrac{1}{2} of \ch{^{31}P} and bounded electron spin-\sfrac{1}{2} play an important role \cite{Kane1998}. The control and readout of the nuclear spin are achieved only through the bound electron spin. Recent papers by Morello \textit{et al.}~\cite{Morello2020} and by McCallum, Johnson, and Botzem~\cite{McCallum2021} present excellent overviews of donor-based qubits.

\begin{figure}
    \centering
    \includegraphics[scale=1]{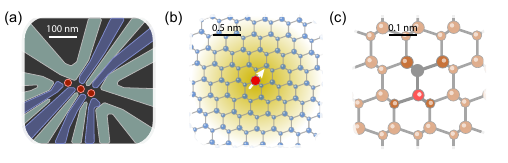}
    \caption[Semiconductor spin qubits]{ Semiconductor spin qubits. (a) An array of three quantum dot spin qubits in silicon (schematically represented as red circles). (b) Phosphor donor in silicon. (c) Negatively charged nitrogen-vacancy center in diamond -- an example of a spin defect.  The figure is adapted from \cite{onizhuk2023thesis}.
}
    \label{fig:semiq}
\end{figure}

Usually, the electronic wavefunctions of shallow defects are significantly delocalized, spanning several nanometers (Fig. \ref{fig:semiq}(b)). The Kohn-Luttinger model is often adopted to describe dopant wavefunctions by treating the donor potential as a perturbation to the average potential of the pristine host crystal (see, for example, \cite{PhysRevB.70.115207}). However, recently, there have been several efforts to characterize the electronic structure of shallow donors completely from first principles~\cite{yan2012, Swift2020, Ma2022ya}.

\subsection{Spin defects}
Point defects are elementary substitutions or vacancies in the crystal lattice of a material. In the case of insulators and semiconductors, a subset of such defects can possess electronic states with energies deep within the bandgap, with localized wavefunctions and well-defined spin states. Some of these defects have both ground and excited state energy levels localized within the band gap of the material, such as nitrogen vacancies in diamond or double vacancies in silicon carbide (SiC). They are known as \textit{color centers} and they absorb light at a frequency smaller than the band gap. To realize a spin-qubit, one in general needs an optically active color center with different spin-states that constitute a two-level system. We also include in this category the rare-earth ions implanted or grown in insulators~\cite{Bertaina2007, Siyushev2014, PhysRevB.105.224106}, where an electronic magnetic moment is associated with the orbital state of a single metal ion.


As an example, consider the most well-studied spin defect for quantum technologies, the negatively charged nitrogen-vacancy center in diamond~\cite{Doherty2013} (Fig. \ref{fig:semiq}(c)). It consists of a nitrogen substituting a carbon atom located next to a vacant lattice site. The ground state of the \ch{NV^-} center is a triplet (\ch{^3A_2}), above which there is an excited triplet state (\ch{^3E}), with two intermediate singlet states (\ch{^1A_1} and \ch{^1E_1})~\cite{PhysRevB.74.104303}. The triplet states have three sub-levels, that differ by the projection $m_s$ of the total electron spin. Two of these sub-levels of the lowest triplet state are used as qubit states. The inter-system crossing rates are spin-state dependent \cite{PhysRevB.86.041202} and by optically exciting the \ch{NV^-} center, one can effectively reset the qubit into the $\ket{m_s=0}$ state within microseconds. Such unique properties of the defect allow one to use the \ch{NV^-} center as a spin qubit even above room temperature~\cite{PhysRevX.2.031001}. Several other defects rose to prominence in recent years, for example, the divacancy in silicon carbide \cite{Koehl2011}, with an electronic structure similar to that of the \ch{NV^-} center, the boron vacancy in hexagonal boron nitride \cite{Gottscholl2021}, and many others.
We refer the reader to the review by G. Wolfowicz \textit{et al.} \cite{Wolfowicz2021} for further details on spin-defect qubits.

From a computational standpoint, the localization of the defect orbitals within a few unit cells of the material allows one to treat the defects in the dilute limit completely from first principles~\cite{RevModPhys.86.253, Dreyer2018, Ivdy2018}, including high-level methods such as the quantum defect embedding theory (QDET)~\cite{Ma2021}. This method treats the states of the defects with quantum chemistry techniques, e.g. configuration interaction, and accounts for the semiconductor environment with density functional theory~\cite{Sheng2022}. Other promising quantum embedding approaches include the density matrix embedding theory (DMET)~\cite{Mitra2021} and, in general, active space-based quantum embedding~\cite{Lau2024}. The \textit{ab initio} investigation of the electronic properties of spin defects remains an active area of research to this day.

\section{Introduction to spin relaxation}\label{sec:spin_relaxation}
\subsection{Spin qubits interacting with the environment}

A qubit is a two-level quantum system characterized by the Pauli matrices $\hat \sigma_x, \hat \sigma_y, \hat \sigma_z$. If the spin of the qubit is $\sfrac{1}{2}$, then the Pauli matrices are equal and physically equivalent to the component of the spin operator $\mathbf{S}=[\hat S_x, \hat S_y, \hat S_z]$ divided by $\hbar/2$. Spin operators for higher spin states are matrices satisfying the same commutation relations as Pauli matrices, but they correspond to irreducible representations of the SU(2) generators \cite{Griffiths_Schroeter_2018} different from those of spin $\sfrac{1}{2}$.

The decoherence of a qubit is intrinsically linked to its environment, and it is determined by both the qubit's location within the host material and the material's intrinsic properties. In general, the environment includes electric and/or magnetic degrees of freedom coupled to those of the qubit. The Hamiltonian of a two-level qubit interacting with its environment can be formally written as \cite{RevModPhys.89.035002}:
\begin{equation}\label{eq:generic_q_hamiltonian}
    \hat H / {\hbar} = \frac{1}{2}\omega \hat \sigma_z + \frac{1}{2}\hat \nu_\parallel \hat \sigma_z + \frac{1}{2}\hat \nu_\perp \hat \sigma_x + \hat H_B,
\end{equation}
where $\omega$ is the energy difference between the qubit levels (qubit frequency), $\hat \nu_\parallel$, and $\hat \nu_\perp$ are operators defined in the Hilbert space of the environment, and $\hat H_B$ is the Hamiltonian describing the interacting environment, not necessarily known a priori. Note that the spin Hamiltonian can contain higher order terms for spins higher than-\sfrac{1}{2} (See Sec. \ref{sec:spinham} for details). The interactions of a qubit with the environment broadly fall within two categories: longitudinal $\hat \nu_\parallel$ interactions (coupled to $\hat \sigma_z$) that modify the qubit frequency, and transverse $\hat \nu_\perp$ interactions (coupled to $\hat \sigma_x$), that may induce a transition between the qubit levels.
Both of these interactions may lead to
the entanglement of the qubit state and the environment, and thus to the degradation of the pure state of the
qubit, even when the environment is in a pure quantum state~\cite{nielsen2000quantum, breuer2002theory}. In solid-state spin systems, the environment is rarely in a pure state and thus the degradation of a qubit can often be relatively well described with semi-classical models. Here, we first describe the generic properties of possible decay processes and provide an intuitive picture of Gaussian classical noise; we then discuss in detail the physical origins of decoherence processes and how to simulate them.

\subsection{Longitudinal relaxation}\label{sec:relaxation}

Transverse interactions $\hat \nu_\perp \hat \sigma_x$ with the environment lead to a change in the diagonal elements of the density matrix, or populations, of the qubit. The change in the populations of qubit energy levels directly leads to a change in the longitudinal magnetization of the qubit $\langle\hat\sigma_z\rangle=\rho_{00} - \rho_{11}$, and its decay is characterized by the relaxation time $T_1$.

One can obtain an intuitive picture of the longitudinal relaxation process by approximating the effect of the environment with that of classical stochastic processes. Hence, the operators on the right hand side of Eq. (\ref{eq:generic_q_hamiltonian}) are replaced by classical time-dependent stochastic variables. In this case, one can relate the relaxation rate $\Gamma_{10}$ to the spectral density $S_p$ of the noise using the Fermi golden rule \cite{PhysRevA.87.022324, PhysRevLett.93.267007}:
\begin{equation}
    \Gamma_{10} = \frac{1}{4} S_p[\omega],
\end{equation}
and in the white noise limit (where the noise auto-correlation time is significantly shorter than the time of the experiment), the qubit dynamics is Markovian, and the population decays exponentially, with a characteristic time equal to the inverse of the relaxation rate obtained from the Fermi golden rule: $T_1 = 1 / \Gamma_{10}$. In most systems of interest, the relaxation is well described by this limit, matching the predictions of classical Bloch equations (Appendix \ref{app:history}).

\subsection{Dephasing}\label{sec:dephasing}

The environmental interactions described by the term $\hat \nu_\parallel \hat \sigma_z$  do not lead to a variation of the population between qubit energy levels. Therefore, in this case all environment-induced dynamical processes are reflected in the changes of the off-diagonal elements of the density matrix $\rho_{10}(t)$, which are physically equivalent to changes in the magnetization in the $xy$-plane, or $\rho_{10}(t)=\frac{1}{2}(\langle{\hat\sigma_x}\rangle + i\langle{\hat\sigma_y}\rangle)$.

The off-diagonal elements of the density matrix are often referred to as \textit{coherences}, as they are proportional to the relative phases of the two superimposed states of the qubit in a given basis. Throughout the text we use $\hat \sigma_z$ basis, as the energy scale is implicitly assumed to be dominated by interactions along the $z$-axis (or, conversely, the $z$-axis is chosen to match the largest energy splitting in the system). Analogous to classical wave sources, the qubit states may be considered coherent if the relative phase of the two states is well-defined. If the phase are completely randomized, the two states are incoherent, and the density matrix represents a classical mixture of the states instead of a quantum superposition. The process during which a qubit loses a well-defined phase between its levels due to the environment is called \textit{decoherence} or \textit{dephasing}.

Tracing out the environmental degrees of freedom, we can write $\rho_{10}(t)$ as:\begin{equation}\label{eq:rho_10_2}
    \rho_{10} (t)  = \rho_{10}(0)\cdot e^{-i\omega t}\cdot \mathcal{L}(t),
\end{equation}
where we define the coherence function $\mathcal{L}(t)$ as:
\begin{equation}\label{eq:coherence_function0}
     \mathcal{L}(t) =  \langle \mathcal{\Tilde{T}}e^{-\frac{i}{2}\int_0^t{\hat \nu_\parallel^{\mathrm{int}} (t)}} \mathcal{T}e^{-\frac{i}{2}\int_0^t{\hat \nu_\parallel^{\mathrm{int}} (t)}}\rangle,
\end{equation}
where $\langle\cdot\rangle$ defines expectation value for the given bath state, $\hat \nu_\parallel^{\mathrm{int}}(t) = e^{i\frac{i}{\hbar}\hat H_B t}\hat \nu_\parallel e^{-\frac{i}{\hbar}\hat H_B t}$ is the noise operator in the interaction frame, $\mathcal{{T}}$ ($\mathcal{\Tilde{T}}$) is the time-ordering (anti-time-ordering) operator, and $\omega$ and $\hat H_B$ are defined in Eq.~\ref{eq:generic_q_hamiltonian}.

Here, we mainly focus on the regime most relevant to solid-state spin qubit technologies, where the dephasing rate significantly exceeds the relaxation rate of the qubit, and thus on models and simulations to compute the coherence function $\mathcal{L}(t)$; in particular, we focus on obtaining $\mathcal{L}(t)$ from first principles calculations and on describing predictive computational frameworks.
The coherence function is usually fitted to a stretched exponential form, $\mathcal{L}(t)=\exp(-\left(\frac{t}{T_2}\right)^n)$ to recover a single phenomenological coherence time $T_2$, describing the typical timescale of the qubit decoherence. Here $T_2$ and $n$ are fitting parameters. Akin to the relaxation process, one can recover $n=1$ only in the limit of white noise. Although it is not always possible to accurately fit $\mathcal{L}(t)$ to the simple formula above, the fit often serves as an easy-to-use estimate to compare results for different systems. We also note that $T_2$ here is a generic parameter for various time constants that one can measure for spin qubit dephasing, as we discuss below.

Specifically, one is concerned with spin dephasing during the qubit's free evolution and under dynamical decoupling pulse sequences. These sequences involve series of $\pi$-pulses applied to extend the coherence time of spin qubits \cite{PhysRevLett.95.180501}.

In the limit of quantum stationary Gaussian noise, we can write the coherence function as an integral over the appropriately symmetrized autocorrelation function of the noise operator \cite{Szakowski2017}:

\begin{equation} \label{eq:gaussnoise}
     L(t) = \exp[-\int_0^t d \tau F(\tau, t)\langle\{\hat \nu_\parallel(\tau)\hat \nu_\parallel(0)\}\rangle],
\end{equation}

where $F(\tau, t)$ is a time-valued filter function, that depends on the application of the dynamical decoupling pulse series\cite{PhysRevA.86.012314},
    $F(\tau, t) = \int_\tau^{2t-\tau}dx y\left(\frac{x + \tau}{2}\right)y\left(\frac{x - \tau}{2}\right)$,
where $y(t)=\pm 1$ is positive or negative after even or odd number of $\pi$-pulses has been applied at time $t$, respectively. Taking the Fourier transform of the right-hand side of the equation, the coherence function is related to the noise spectral density multiplied by a frequency filter function $F(\omega, t)$:
\begin{equation} \label{eq:sd_gn_dd}
     \mathcal{L}(t) =\exp[-\frac{1}{4\pi}\int_{-\infty}^\infty{S_p[\omega]F(\omega,t)}].
\end{equation}

where the noise spectral density is directly related to the correlation function of the noise. In the classical limit, we have \cite{RevModPhys.82.1155}:
\begin{equation}\label{eq:somega}
    S_p[\omega]= \int_{-\infty}^{\infty} \langle\nu_\parallel(\tau)\nu_\parallel(0)\rangle e^{i\omega \tau} d\tau.
\end{equation}.

\begin{figure}
    \centering
    \includegraphics[scale=1]{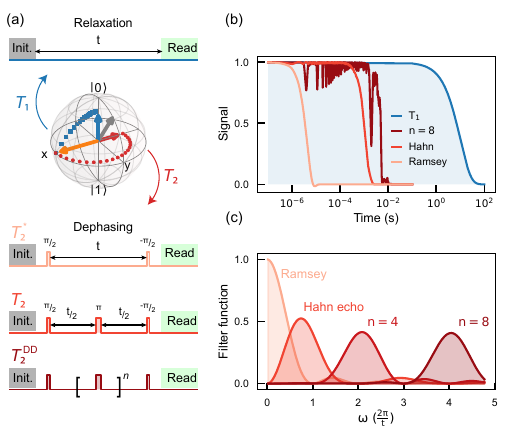}
    \caption{Decoherence of a qubit. (a) Different decay processes of a single qubit and pulse sequences used to measure corresponding coherence times. Top: Relaxometry; bottom: Ramsey, Hahn-echo, and generic dynamical-decoupling sequences. (b) Examples of the signal obtained in experiments, using simulated data for the \ch{NV^-} center as an example. (c) Filter functions for different dephasing experiments (see text).}
    \label{fig:introcoh}
\end{figure}

Usually, in experiments, the transverse magnetization of a single spin qubit cannot be measured directly. Instead, the coherence is measured by mapping it onto the population of the qubit and reading out the population signal (Fig. \ref{fig:introcoh}(a)).

The first important pulse sequence used in many experiments is the Ramsey protocol. This sequence interrogates a qubit with $\frac{\pi}{2}$ pulses: the first prepares a superposition state, and the second maps the $\sigma_x$ expectation value on the population of the $\ket{0}$ state \cite{hong2013}. Thus a
Ramsey sequence is the proxy for probing the free evolution of the qubit (known as free induction decay in magnetic resonance). The decay of the spin coherence under free evolution averaged over a temporal ensemble (many repetitions of the measurement for a single qubit) or over a spatial ensemble (many measurements for different spins of the same type), is characterized by the so-called inhomogeneous coherence time $T_2^*$.

The inhomogeneous coherence time $T_2^*$ is limited by the quasi-static noise present in the experiment, i.e. by the noise that is  constant for each measurement but varies from measurement to measurement. In spin qubit systems, the static noise is dominated by the fluctuations of the Overhauser field (magnetic field induced by other spins in the systems).
The autocorrelation of such noise is trivial, $\langle\nu_\parallel(\tau)\nu(0)_\parallel\rangle=\langle\nu(0)_\parallel\nu(0)_\parallel\rangle=\langle\nu^2\rangle$, and for a single spin we obtain the following equation for the decay:
\begin{equation}
    L_\text{static}(t)=e^{-\frac{\langle\nu^2\rangle}{2}{t^2}}.
\end{equation}

Note that one expects a Gaussian decay for a single qubit, but an exponential decay for the spatial ensemble of spins \cite{PhysRevB.77.245212}.

When using a Hahn-echo sequence of pulses, one adds a $\pi$ pulse in the middle of the Ramsey experiment that effectively removes the effect of any static variations of the qubit frequency on the final signal \cite{PhysRev.80.580}. The characteristic time measured in the Hahn-echo experiment is usually called $T_2$. However, in the spin qubit literature, any coherence time obtained with dynamical decoupling protocols where additional $\pi$ pulses are applied during the evolution time \cite{PhysRevB.85.155204}) have been referred to as $T_2$; the reader needs to be cautious not to confuse the coherence times obtained with different pulse sequences.

To explain the extended coherence time under dynamical decoupling, we can adopt filter-functions formalism. The Hahn echo filter function is equal to $F_\text{HE}(\omega,t)=\frac{8\sin^4(\frac{\omega t}{4})}{\omega^2}$ \cite{PhysRevB.77.174509}. The filter functions for Ramsey, Hahn-echo, and other experiments are plotted in Fig. \ref{fig:introcoh}(c). Under dynamical decoupling, the filter functions approach zero at low noise frequencies; thus, applying periodic $\pi$-pulses allows one to remove the low-frequency components of the noise and, as a consequence, significantly prolong the coherence time as long as the noise is sufficiently "slow", namely the noise spectrum decays at large frequencies. This decay is typically accompanied by the super-exponential decay of the coherence signal, $\mathcal{L}=\exp(-(t/T_2)^a)$ with $a>1$. A comparison between the typical timescales probed with spin qubit measurements is given in Figure \ref{fig:introcoh}(b).

Note that the presence of transverse noise can also be a source of decoherence; in the presence of white transverse noise, the characteristic time under any dynamical decoupling never exceeds twice the relaxation time, $T_2 \leq 2T_1$ \cite{nielsen2000quantum}. To distinguish between relaxation-induced decoherence and so-called ``pure" dephasing, the time $T_\phi$ is used to denote pure dephasing contributions. The total coherence time is then computed as $\frac{1}{T_2}=\frac{1}{2T_1}+\frac{1}{T_\phi}$ \cite{chirolli2008}. In the systems of interest to this Colloquium $T_2 \ll T_1$, and the contribution of the relaxation to dephasing is usually omitted.

The representation of the environment of spin qubits adopted in this section relies on several major assumptions, including that a classical stochastic process can faithfully reproduce the dynamics of the environment and that the noise is stationary and Gaussian in nature. The main challenge in identifying a classical noise source describing experimental results lies in the so-called ``backaction" of the qubit \cite{Yang_2016}, namely the fact that the state of the environment depends on its interactions with the qubit. Interestingly, the backaction of the qubit represents an opportunity to define a non-trivial sensing modality enabling the recovery of the properties of many-body quantum systems \cite{Wang2021}.
Only in the case of a limited class of experimentally accessible systems can this backaction be neglected, and a classical noise source faithfully reproduces quantum-mechanical predictions \cite{PhysRevB.92.161403}. For the majority of the spin-limited decoherence processes, one needs to consider a full quantum-mechanical evolution of the magnetic environment surrounding a qubit, which motivates the development of accurate numerical models of spin decoherence.

\section{First-principles relaxation models}\label{sec:theory}
Prediction of the spin qubits dynamics in solids from first principles consists of two major steps:

\begin{itemize}
    \item The construction of an effective Hamiltonian representing the interaction of the qubit with the environment using a first principles approach, e.g., density functional theory.
    \item The determination of the qubit dynamical properties using the effective Hamiltonian and appropriate numerical approximations.
\end{itemize}

Similar to early magnetic resonance experiments (Appendix \ref{app:history}), the main media of interest to understand the dynamics of a spin qubit include magnetic environments (consisting of other spin-carrying particles) and the vibrating nuclei of the host material. However, there may be additional interactions determining the qubit dynamical properties; for example, the interactions with electric field originating from charges in the solid can induce both relaxation and dephasing, as described in the recent theoretical work by Candido and Flatt\'e \cite{candido2023interplay}, as well as in experimental studies \cite{PhysRevB.81.075214, PhysRevB.102.165427, PhysRevX.9.031052}. Identifying and understanding the effect of all factors that influence the spin qubit dynamics, beyond magnetic noise and lattice vibrations, is an active area of research. At present the interactions best characterized with ab initio approaches are those leading to spin-spin and spin-phonon relaxation pathways.

\begin{figure}
    \centering
    \includegraphics[scale=1]{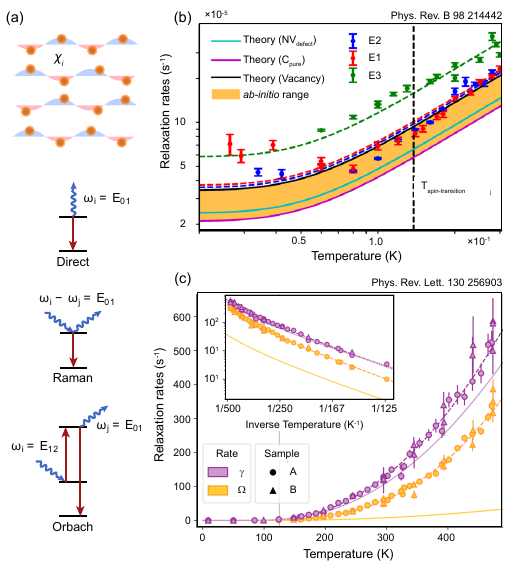}
    \caption{Spin-phonon relaxation. (a) Schematic representation of lattice vibrations and main types of relaxation processes, top to bottom: Direct, Raman, Orbach. (b) The measured spin-lattice relaxation rates for three different samples of NV in diamond (E1, E2, and E3; data taken from \cite{Astner2018}). Dashed lines are fits to the data. The theoretical results (yellow range) depend on the phononic density of states. Adapted from \cite{PhysRevB.98.214442} under CC 4.0 license. (c) Temperature dependence of relaxation rates $\Omega$ (population decay between $\ket{0}\leftrightarrow\ket{\pm 1}$ levels) and $\gamma$ ($\ket{+1}\leftrightarrow\ket{- 1}$). Darker lines are fits to the data. Lighter solid lines show relaxation rates predicted by \textit{ab initio} calculations. Inset: semilog plot of relaxation rates versus inverse temperature. Adapted from \cite{PhysRevLett.130.256903} .}
    \label{fig:phonons}
\end{figure}

\subsection{Spin-phonon relaxation}\label{sec:phonons}

At sub-Kelvin temperatures and higher, the relaxation of the spin qubits is almost entirely dominated by the interactions between spin degrees of freedom and lattice vibrations. In general, we can write the Hamiltonian describing the interactions between the central spin and phonons as:
\begin{equation}\label{eq:spinphonon}
    \hat H_\mathrm{s{\text -}ph} = \sum_i \frac{\partial \hat H_S}{\partial \chi_i}\hat\chi_i + \sum_{i,j} \frac{\partial^2 \hat H_S}{\partial \chi_i\partial \chi_j}\hat\chi_i\hat\chi_j,
\end{equation}
where $\hat H_S$ is the central spin Hamiltonian (described in detail later in the text), $\hat\chi_i=\hat{b}_i+\hat{b}^\dagger_i$ are generalized nuclear positions, $\hat{b}_i$ and $\hat{b}_i^\dagger$ are annihilation and creation operators of phonon bosonic modes. Several theories have been proposed for extended electronic states in bulk materials \cite{PhysRevB.106.174404, Xu2020}, which yield a good agreement with experiments for computed spin relaxation times \cite{PhysRevLett.129.197201, Xu2024};
however, describing the spin relaxation of localized electronic states associated with impurities using ab initio methods remains an open challenge in the field.

One possible approach is to treat the phonon environment as a fast bath, where the interactions between spin and lattice vibrations do not affect the state of the phonon bath. In this case, one may utilize perturbative treatments to recover relaxation rates from first principles.

At sub-Kelvin temperatures, the relaxation of the spin qubit is dominated by single-phonon processes, with the energy of the phonon matching the frequency of the qubit (Fig. \ref{fig:phonons}(a)). At these ultra-low temperatures, the rates obtained by using the Fermi golden rule and first-principles calculations of the phonon spectrum at the DFT level of theory show good agreement with experiments for the \ch{NV^-} center in diamond \cite{Astner2018, PhysRevB.98.214442} (Fig. \ref{fig:phonons}(b)). One can also use single-phonon processes to estimate the effect of the phononic bath on the dephasing of the spin qubit \cite{simoni2022phonon}.

Two-phonon processes (Fig. \ref{fig:phonons}(a)) play a dominant role at higher temperatures. Starting with the general Hamiltonian (Eq. \ref{eq:spinphonon}), we can identify two different origins of the two-phonon relaxation pathways: either a second-order transition due to the linear term of the Hamiltonian (Orbach process) or a first-order contribution due to the quadratic term (Raman scattering). As well known, Raman scattering involves a transition to a virtual state, while Orbach processes usually involve a low-lying state.

Combined models to simultaneously capture the effect of single- and two-phonon processes have been proposed by Norabuena \textit{et al.}~\cite{PhysRevB.97.094304} using the Fermi golden rule at various orders of perturbation theory and by Lunghi \textit{et al.}~\cite{lunghi2022} using master equation approaches. The recent works by Cambria \textit{et al.} and Mondal and Lunghi \cite{PhysRevLett.130.256903, Mondal2023} show agreement between results obtained with higher order perturbation theory and the experimental data for spin defects in diamond and hBN, in a wide range of temperatures (Fig. \ref{fig:phonons}(c)), thus showing promise for the predictive power of first-principles simulations of spin-phonon relaxation processes.

We note that all existing first-principles predictions of spin-phonon relaxation processes assume an intrinsic separation between the time scales of spin-qubit relaxation and phonon thermalization. This assumption allows for the application of the perturbative approaches previously described. However, this separation is not a valid approximation in the presence of magnetic noise, namely when the qubit interacts with a magnetic environment, as discussed in the remainder of this Colloquium.


\subsection{Spin-spin relaxation}\label{sec:spins}
In many systems of interest where one can isolate and control a single spin for quantum information processing, there is usually a mismatch between the characteristic excitation frequency of the target spin and that of the environmental spins. Therefore, the energy exchange between the central spin and its spin environment is significantly suppressed. In this common case, the effect of the spin environment on the qubit population dynamics can be safely neglected. However, one can engineer the environment to amplify the spin exchange and, for example, observe emergent macroscopic hydrodynamics phenomena \cite{Zu2021}.

We refer the reader interested in estimating the effect of the magnetic environment on the qubit relaxation from first principles, to a series of papers by Iv\'ady and coworkers \cite{PhysRevB.101.155203, BulanceaLindvall2021, PhysRevApplied.19.064046}. In these works, the authors introduce a cluster approach to estimate the population dynamics of the central spin in a spin bath; they apply their method to gain insight into the spin bath-limited relaxation time of spin defects in isotopically purified materials. Here, we focus on the opposite regime of pure dephasing dynamics.

\subsection{First principles simulations of the spin Hamiltonian}\label{sec:spinham}
\begin{figure}
    \centering
    \includegraphics[scale=1]{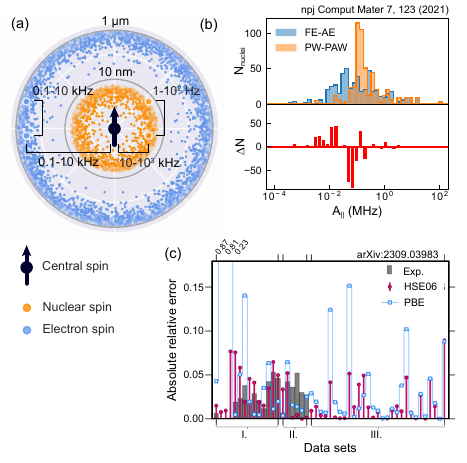}
    \caption{Spin environment of solid-state spin qubits. (a) Schematic representation of the spin baths surrounding an \ch{NV^-} center in diamond, plotted on a semi-logarithmic scale. Numbers show typical coupling values between central spin and electron, nuclear spin baths, as well as typical interaction strength within the bath given natural concentration of nuclear spins 1.1\% \ch{^{13}C} in diamond and common concentration of paramagnetic impurities of $10^{15}$ cm$^{-3}$. (b) Histogram of computed $A_\parallel$ components of the hyperfine tensor, obtained with pseudopotential (PW-PAW) and all-electron (FE-AE) methods, with a 64-atom diamond supercell with an \ch{NV^-} center. $\Delta N$ is the difference in the number of nuclei with given hyperfine parameters computed with the two methods. Adapted from \cite{Ghosh2021} under CC 4.0 license. (c) Absolute relative error (ARE) on computed values of  $A_\parallel$ calculated with pseudopotentials and plane waves basis set and improved integration methods. Gray columns depict the ARE of the experimental data, red thin columns with circles depict the ARE of the computed hyperfine values obtained with the hybrid HSE06 functional \cite{hse06}, and light blue line with squares depict the ARE of the theoretical values obtained with the PBE functional \cite{PhysRevLett.77.3865}. The values provided on the upper horizontal axis represent the PBE absolute relative errors that are out of the range of the vertical axis. Adapted from \cite{takacs2023accurate} .}
    \label{fig:spins}
\end{figure}

Two types of spin baths contribute to the effective spin qubit environment (Fig. \ref{fig:spins}(a)): the nuclear and electron spin baths, which have different effective magnetic moments and spatial concentrations. The nuclear spins that significantly affect the central spin's evolution are typically located within a few nanometers, while the bath electron spins are within distances up to a few microns.

A nuclear spin bath is an intrinsic characteristic of the host material of the spin defect. It is usually determined by the natural concentration of spinful isotopes of the elements constituting the crystal lattice. For example, natural diamond contains 98.9\% of the \ch{^{12}C} isotope with spin zero, and 1.1\% of the \ch{^{13}C} nuclei with spin-$\sfrac{1}{2}$. Therefore, the nuclear spin bath of diamond is by far dominated by \ch{^{13}C} nuclei. One can isotopically purify the diamond host to obtain a virtually nuclear spin-free environment \cite{Itoh2014} and enhance the coherence of spin qubits \cite{Balasubramanian2009, BarGill2013}, thus improving their efficiency in quantum network applications \cite{Bradley2022}. 

However, obtaining isotopically purified hosts is not always possible or easy. For example, hexagonal boron nitride (hBN), a promising 2D platform for quantum technologies \cite{Aharonovich2022}, contains close to 100\% concentration of spinful isotopes of both nitrogen and boron, making the nuclear spin bath a dominant  source of decoherence in any regime \cite{Ramsay2023, Rizzato2023, Guo2023, gao2023nanotube}.

The electron spin bath consists of spinful electron spin- defects in the bulk or surface of the host material. These defects may be the same or different from the central spin. Usually, unwanted defects are introduced during the growth of the material and are almost unavoidable as their presence may be required to create the desired conditions to stabilize the spin qubit.
For example, consider again the negatively charged \ch{NV^-} center in diamond. To grow \ch{NV^-} centers, one needs to have a sufficiently high concentration of nitrogen in diamond and at the same time introduce some vacancies into the lattice. These nitrogen precursors possess an electron magnetic moment acting as a major noise source in diamond \cite{PhysRevB.102.134210, BarGill2012}.

To study the dynamics of spin systems, one would usually utilize a model Hamiltonian in the Hilbert space of central and bath spins, which contains spin variables describing a central spin $\mathbf{S}$ and the environmental spins $\mathbf{I}$. 
The spin interactions are expressed as products of spin operators with the external magnetic field and coupling constants (parameters). These coupling parameters implicitly contain information about the many-electron wavefunction, geometry, and other structural parameters of the spin and host materials. 

For example, consider an electron with spin-1 in an external magnetic field and in the presence of a nuclear spin bath. For this system, the spin Hamiltonian can be separated into the following terms:
\begin{equation} \label{eq:totalham}
    \hat H= \hat H_\mathrm{S} + \hat H_\mathrm{SB} + \hat H_\mathrm{B},
\end{equation}
where $\hat H_\mathrm{S} = \hat H_\mathrm{Z} + \hat H_\mathrm{ZFS}$ is the central spin Hamiltonian, which for an electron spin-1 in a nuclear bath includes the Zeeman term $\hat H_\mathrm{Z}$ and zero-field splitting $\hat H_\mathrm{ZFS}$, $\hat H_\mathrm{SB}$ is the spin-bath (B) Hamiltonian and $\hat H_\mathrm{B} = \sum_n (\hat H^n_\mathrm{Z} + \hat H^n_\mathrm{Q} )+ \hat H_\mathrm{BB}$ is the nuclear spin Hamiltonian including the Zeeman terms $\hat H^n_\mathrm{Z}$, quadrupole interactions $\hat H^n_\mathrm{Q}$, where the sum over $n$ extends over all bath spins, and intra-bath interactions described by $\hat H_\mathrm{BB}$.
Below, we discuss how to compute the parameters of the spin Hamiltonian. 

We begin with the Zeeman splitting. The magnetic field-electron spin interaction can be expressed as:
\begin{equation}\label{eq:gtensor}
    \hat H_\mathrm{Z}=\frac{\mu_B}{\hbar} \mathbf{BgS},
\end{equation}
where $\mathbf{S}=(\hat{S}_x, \hat{S}_y, \hat{S}_z)$ is a vector of electron spin operators, $\mu_B$ is the Bohr magneton, equal to the magnetic moment of an electron, $\mathbf{B}=(B_x, B_y, B_z)$ is an external magnetic field, and $\mathbf{g}$ is a tensor, describing the interactions between the central spin and a magnetic field.
The $\mathbf{g}$ tensor contains a contribution from the spin-field interactions and orbital momentum-field interactions.
The components of the $\mathbf{g}$ tensor can be written as:
\begin{equation}
    g_{ab}=g_e\delta_{ab}+\Delta g^\mathrm{RMC} \delta_{ab} + \Delta g_{ab}^\mathrm{GC} ,
\end{equation}
where $g_e \approx 2.002318$ is the free electron value (dominating contribution), $\Delta g^\mathrm{RMC}$ is the relativistic mass correction, and the term $\Delta g_{rs}^\mathrm{GC}$ is a diamagnetic correction. Higher order terms arise from orbital Zeeman and spin-orbital interactions \cite{Neese2001}. 
The free electron spin gyromagnetic ratio $\gamma_e$ is connected to $g_e$ as $\gamma_e=g_e\frac{\mu_B}{\hbar}$.
In the case of nuclear spins, the interactions are usually written as follows:
\begin{equation}  
\hat H^n_\mathrm{Z} = \gamma_n[1 + \mathbf{\sigma}_n] \mathbf{B}\mathbf{I},
\end{equation}
where $\mathbf{I}=(\hat{I}_x, \hat{I}_y, \hat{I}_z)$ are nuclear spin operators, and $\mathbf{\sigma}_n$ is a chemical shift tensor, describing the local change in the magnetic field due to the electronic structure \cite{chem_shift} of the solid, and $\gamma_n$ is nuclear gyromagnetic ratio. In the systems of interest to this Colloquium, chemical shifts are usually small relative to the total Zeeman interaction of the order of a few parts per million (ppm), and they can be safely neglected in the simulations of spin qubits in the solid state.

Next, we consider the interaction terms in spin-1 systems and higher. 
For example, the triplet electronic state of a spin defect that gives rise to a spin-1 system includes two unpaired electrons. In the effective spin Hamiltonian, the interactions between two unpaired electrons are included in one term known as zero field splitting, expressed as:
\begin{equation}\label{ZFS}
    \hat H_\mathrm{ZFS}=\mathbf{SDS},
\end{equation}
The tensor \textbf{D} completely describes the interactions between the electron spins. To first order \textbf{D} is given by spin dipole-dipole interactions and its elements are~\cite{PhysRevB.77.035119}:
\begin{widetext}
\begin{equation}
    D_{ab}=\frac{1}{2S}\frac{\mu_B}{4\pi}\gamma_e^2\hbar^2\int_{-\infty}^{+\infty}\int_{-\infty}^{+\infty}{\frac{\abs{\mathbf{r}_1 - \mathbf{r}_2}^2\delta_{ab} - 3 (\mathbf{r}_1 - \mathbf{r}_2)_a (\mathbf{r}_1 - \mathbf{r}_2)_b}{\abs{\mathbf{r}_1\mathbf{r}_2}^5}\rho_s(\mathbf{r}_1,\mathbf{r}_2)d\mathbf{r}_1 d\mathbf{r}_2},
\end{equation}
\end{widetext}
where $\rho_s(\mathbf{r}_1,\mathbf{r}_2)$ is a two-particle density.
At second order, spin-orbit coupling contributes to \textbf{D}~\cite{neese2006importance, Neese2007, Duboc2010}. One can diagonalize the \textbf{D} tensor and rewrite the interaction as:
\begin{equation}\label{eq:ZFS2}
    \hat H_\mathrm{ZFS}=D(\hat S_z^2 - \frac{1}{3} S(S+ 1)) + E (\hat S_x^2 - S_y^2),
\end{equation}
The scalar $D$ is known as axial (parallel) ZFS and the scalar $E$ as transverse ZFS.

For larger numbers of unpaired electrons, one may need to include higher-order operators in the Hamiltonian, usually represented as a set of Stevens operators (see, for example, \cite{stoll2006}).

The nuclear spin Hamiltonian contains terms that describe the orientation dependence of the electronic interactions of the nuclei. Nuclei do not possess an electric dipole moment \cite{levitt2008spin}, and their electric multipoles are equal to zero starting from $2I$.
Therefore, for $I=\sfrac{1}{2}$, no additional electronic interactions appear between nuclei and the electronic spin. But for any higher spins, there are interactions of quadrupolar order or higher. 

The quadrupole interactions with the electronic field are given by:
\begin{equation} \label{eq:quadrupole}
    \hat H^n_\mathrm{Q} = \frac{eQ}{2I(2I-1)} \mathbf{I}_n \mathbf{V}\mathbf{I}_n ,
\end{equation}
where the components of $\mathbf{V}$ are the second derivative of the electrostatic potential at the position of the nucleus:
\begin{equation}
    V_{ab} = \nabla_a\nabla_b\left[-\int_{-\infty}^{+\infty}{d\mathbf{r'}\frac{n(\mathbf{r'})}{|\mathbf{r-r'}|}} + \sum_{i \ne n}\frac{Z_i}{|\mathbf{r}_i|} \right]_{\mathbf{r}=0},
\end{equation}
$e$ is the electronic charge and $Q$ is the nuclear quadrupole moment, $Z_i$ is the charge of the $i$-th nucleus, $n(\mathbf{r})$ is the electronic density.

Interactions between the central spin and the $n$-th bath spins can be collected into a single term:
\begin{equation}
    \hat H_\mathrm{SB}=\sum_n \mathbf{S}\mathbf{A}_n\mathbf{I}_n,
\end{equation}
where we use $\mathbf{I}_n$ for both electron and nuclear bath spin operators.
In the case of an electron central spin and the nuclear bath spins, $\mathbf{A}$ is known as hyperfine coupling, which should be calculated for each nuclear spin. It consists of three different components: isotropic (known as Fermi contact term), anisotropic (dipolar term), and spin-orbital coupling term:
\begin{equation}\label{HF tensor}
    \mathbf{A}=\mathbf{A}^\mathrm{iso}+\mathbf{A}^\mathrm{dip}+\mathbf{A}^\mathrm{SO},
\end{equation}
where the tensor components of the first two terms are given by:
\begin{equation}
    A_{ab}^\mathrm{iso}=-\delta_{ab}\frac{1}{3S}\mu_0\gamma_e\gamma_n\hbar^2n_s(\mathbf{R}_n)
\end{equation}
\begin{equation}\label{eq:adipole}
    A_{ab}^\mathrm{dipole} = \frac{1}{2S}\frac{\mu_0}{4\pi}\gamma_e\gamma_n\hbar^2\int_{-\infty}^{+\infty}{\frac{\abs{\mathbf{r}}^2\delta_{ab} - 3 \mathbf{r}_a \mathbf{r}_b}{\abs{\mathbf{r}}^5}n_s(\mathbf{r})d\mathbf{r}},
\end{equation}
The quantities $\gamma_e$ and $\gamma_n$ are gyromagnetic ratios of the central spin and the nuclear spin, respectively, $n_s$ is the spin density, $\mathbf{r}$ is the distance of the qubit from the nuclei \cite{PhysRevMaterials.3.043801}.
Higher-order contributions $\mathbf{A}^\mathrm{SO}$ arise from spin-orbital interactions. In experiments, one usually measures two main components of the hyperfine tensor, the parallel hyperfine $A_\parallel=A_{zz}$, and perpendicular hyperfine $A_\perp=\sqrt{A_{xz}^2+A_{yz}^2}$ couplings.

In the case of electron bath spins, one can usually approximate the $\mathbf{A}$ tensor using only dipole-dipole interactions (Eq. (\ref{eq:adipole})). In a sparse electronic bath, where the average distance between electron spins is significantly larger than the spread of the square moduli of the electronic wavefunctions, we can treat each electron spin as a magnetic point dipole, and then the interaction term further simplifies to:
\begin{equation}\label{eq:apoint}
    A_{ab}^\mathrm{dipole} =\frac{\mu_0}{4\pi}\gamma_e\gamma_n\hbar^2{\frac{\abs{\mathbf{r}}^2\delta_{ab} - 3 \mathbf{r}_a \mathbf{r}_b}{\abs{\mathbf{r}}^5}}.
\end{equation}

The interactions between bath spins can be expressed in a similar fashion:
\begin{equation}
    \hat H_\mathrm{BB}=\mathbf{I}_i\mathbf{J}_{ij}\mathbf{I}_{j},
\end{equation}
where $\mathbf{J_{ij}}$ has two main components: dipolar coupling (for both electron and nuclear spin baths) and J-coupling (nuclear spin bath only). In a sparse spin bath, one usually treats both nuclear and electron spins as point dipoles and the components of the respective tensors are computed using equation (\ref{eq:apoint}). The J-coupling term represents the indirect interactions of nuclei through the electron cloud. This interaction requires two nuclei to be connected through a small number of chemical bonds, and thus decays rapidly with distance \cite{levitt2008spin}.

The availability of efficient and accurate approaches to numerically compute the terms entering the spin Hamiltonian is an essential prerequisite for robust predictions of the spin qubit dynamics. 
One may adopt several levels of theory to approximate the coupling between the central spin and the bath. In systems where the orbitals of the electron spin are fairly localized (such as spin defects in solids) and the spin bath consists of very sparse nuclear spins, the point-dipole approximation (Eq. (\ref{eq:apoint})) may suffice. In systems where the wavefunctions of the electron spin is delocalized (such as shallow defects or quantum dots), one may need to use models to approximate the electron wavefunction and estimate the spin density (e.g., using the $k\cdot p$ method \cite{PhysRevB.68.115322, PhysRevB.74.035322}).

Once a level of theory is adopted, it is important to verify the robustness of the chosen numerical approaches. For example, Ghosh \textit{et al.}~\cite{Ghosh2021} showed how different methods to compute the spin couplings within DFT (using either pseudopotentials or all-electron calculations) may lead to qualitatively different distributions of the hyperfine parameters for a given defect in a supercell (Fig.~\ref{fig:spins}(b)). The authors emphasized how the numerical treatment of the interaction of the electronic wavefunction in the proximity of the nuclei may result in significantly different predictions of the coherence times.
Another interesting example is given in the paper by Tak\'acs and Iv\'ady~\cite{takacs2023accurate}, where the authors described the importance of finite-size scaling to obtain accurate DFT-predictions of the hyperfine couplings, when adopting a supercell approach (Fig. \ref{fig:spins} (a)). 

Regarding the predictions of values of the ZFS, theoretical calculations in the solid state have achieved excellent accuracy \cite{Bodrog2013, PhysRevB.97.115135}, also in describing trends, as a function of distance for the ZFS of defects in proximity of surfaces \cite{Zhu2023}; in addition, computed values of the ZFS may be useful in the identification of new spin qubits \cite{Davidsson2019}.

For additional information on DFT investigations of the electronic properties of spin defects, we refer the reader to two reviews by Gali~\cite{gali2023recent} and Iv\'ady~\cite{Ivdy2018}.

\subsection{Cluster expansion methods for coherence}
If the bath size is relatively small (up to tens of spins), the coherence of the central spin can be directly inferred from the evolution of the whole bath. Such evolution can be obtained by the exact diagonalization of the Hamiltonian (e.g. \cite{PhysRevLett.102.237601}), or using approximate methods. For example, one can use the time-dependent Density Matrix Renormalization Group (tDMRG) if one is interested in the dynamics of a particular bath state \cite{PhysRevLett.93.040502, PhysRevB.77.205419}, or Dynamical Mean Field Theory (DMFT) \cite{PhysRevResearch.5.043191} which provides a semiclassical description of the bath evolution.

In many systems of interest, however, straightforward simulations of the central spin dynamics remain numerically prohibitive. The leading numerical approach to solve for the quantum dynamics of the central spin consists in factorizing the effect of the environment on the coherence function $L$ into a product of cluster contributions. For a detailed overview of the cluster expansion techniques adopted in dynamical simulations and for a theoretical description of the noise models,  we recommend the review by Yang, Ma, and Liu~\cite{Yang_2016}. Here, we focus on the application of cluster expansion techniques to quantitative predictions of coherence times of spin qubits in solids and molecular systems, after briefly introducing the numerical techniques.

The cluster expansion methodology was originally developed using just a pairwise factorization of the spin decoherence. In particular, Witzel, de Sousa, and Das Sarma~\cite{PhysRevB.72.161306}  and Yao, Liu, and Sham~\cite{PhysRevB.74.195301} proposed approaches where only the interactions of pairs of spins in the bath were considered, with a convenient mapping of the spin pair dynamics onto non-interacting pseudo-spins presented in Ref.~\cite{PhysRevB.74.195301}. A few years later, Maze, Taylor, and Lukin~\cite{PhysRevB.78.094303} used a disjointed cluster approximation, in which they treated the bath of the central spin (an \ch{NV^-} center) as a set of completely disconnected clusters, and the evolution of these clusters was computed separately. Saikin, Yao, and Sham~\cite{PhysRevB.75.125314} introduced a linked cluster expansion (LCE) to obtain the coherence function of a central spin in the spin bath,  by rewriting the function as an exponent of the sum of \textit{linked} diagrams $\hat \pi$ (Eq.~(\ref{eq:coherence_function0})):

\begin{align}\label{seq:iint}
&\mathcal{L}(t)=\langle\mathcal{T} e ^{\int ^{t }_{0} \hat \nu_\parallel^{\mathrm{int}} (t') dt' }\rangle = e ^{\langle\hat \pi\rangle} 
, \\
\label{seq:lce.pi}
& \hat \pi = \sum_{n=0}^\infty \int ^{t }_{0} \! dt _{1}
\ldots \int ^{t _{n-1} }_{0} \! dt _{n} \langle \! \langle 
\hat \nu_\parallel^{\mathrm{int}}  (t_1) \ldots 
\hat \nu_\parallel^{\mathrm{int}}  (t_n) \rangle \! \rangle 
,
\end{align}
where $\mathcal{T} $ is the time-ordering operator, and $\langle \! \langle \cdot \rangle \! \rangle $ are fully connected diagrams obtained by applying Wick's theorem on a series expansion of the left-hand side of Eq.~(\ref{seq:iint}). By truncating the sum at a given order $n$ of the diagrams, one can obtain a solvable model to numerically approximate the coherence function. The LCE is a useful tool to quantify the impact of the correlations of clusters of different sizes in the system \cite{Ma2014}. In principle, the LCE method allows one to solve for the spin dynamics under any conditions. In practice, however, evaluating the contractions of higher correlations is a challenging numerical problem \cite{Yang_2016}. 

Finally, Yang and Liu \cite{PhysRevB.78.085315, PhysRevB.79.115320} introduced the cluster-correlation expansion (CCE) method that combines all previous ideas mentioned above under the same umbrella.

The core concept of the CCE approach is that the coherence function can be factorized into a set of irreducible contributions from overlapping clusters of bath spins of various sizes:
\begin{equation}\label{eq:l_cce}
    \mathcal{L}(t) = \prod_{C} \Tilde{L}_C(t) = \prod_{i}\Tilde{L}_{\{i\}}(t)\prod_{i,j}\Tilde{L}_{\{ij\}}(t)...,
\end{equation}
where $\Tilde{L}_{\{i\}}(t)$ is the contribution of the single bath spin $i$ and $\Tilde{L}_{\{ij\}}(t)$ is the irreducible contribution of the spin pair $i,j$ and so on. The maximum size of the cluster $n$ is equal to the order of the CCE approximation. Each cluster contribution is defined recursively as:
\begin{equation}\label{eq:l_contribution}
    \Tilde{L}_C = \frac{L_{C}}{\prod_{C'}\Tilde{L}_{C'\subset C}},
\end{equation}
where in the equation above the time dependence is implied; $L_{C}$ is the coherence function of the qubit, including only interactions with the bath spins in a given cluster $C$, and $\Tilde{L}_{C'}$ are contributions of the sub-cluster $C'$ of $C$. 

In the original formulation of the CCE approach, the total Hamiltonian of the system (Eq.~(\ref{eq:totalham})) is projected onto a sum of two effective Hamiltonians:
\begin{equation}
    \hat H = \ket{0}\bra{0}\otimes\hat H^{(0)} + \ket{1}\bra{1}\otimes\hat H^{(1)},
\end{equation}
where $\hat H^{(\alpha)}$ with $\alpha=0,1$ is an effective Hamiltonian acting on the bath when the central spin is in the qubit $\ket{\alpha}$ state:
\begin{equation}\label{eq:projectedham}
    \hat H^{(\alpha)} = E_\alpha + \bra{\alpha}\hat H_\mathrm{SB} \ket{\alpha} + \hat H_\mathrm{B} + \hat H^{(\alpha)}_\mathrm{PT},
\end{equation}
and $\hat H^{(\alpha)}_\mathrm{PT}$ are terms originating from higher orders of  perturbation theory \cite{pycce}.

The coherence function of the qubit interacting with the cluster $C$ can be computed as:
\begin{equation}
    L_{C}(t) = \Tr[\hat U_C^{(0)}(t)\hat \rho_C \hat U_C^{(1) \dagger}(t)],
\end{equation}
where $\hat U_C^{(\alpha)}(t)$ is the time propagator defined in terms of the effective cluster Hamiltonian $\hat H_C^{(\alpha)}$ and the pulse sequence of the experiment one wishes to model (Fig. \ref{fig:introcoh}). The cluster Hamiltonian $\hat H_C^{(\alpha)}$ has the same expression as the total effective Hamiltonian (\ref{eq:projectedham}) but it only includes the spins inside the cluster $C$. 

In principle, in the CCE method one performs a nontrivial summation of linked cluster diagrams, i.e.  CCE calculations of order $n$ do not only include all LCE clusters of up to order $n$, but \textit{also} include all the $m$ higher-order diagrams (with $m>n$) involving up to $n$ distinct spins~\cite{PhysRevB.78.085315}. In practice, the additional resummation over high-order diagrams performed in CCE calculations leads to improved convergence. However, we emphasize that the main advantage of the CCE method lies in a significantly simpler definition of cluster contributions compared to all other methods previously proposed in the literature. This definition makes the software implementation of the cluster expansion trivial and expandable to any order, for any pulse sequence of interest. The computational simplicity of the CCE method led to an explosion of new predictions of the spin dynamics in various regimes, as we will discuss in Sec. \ref{sec:applications}.

We note that the cluster expansion method is only well-defined and can be analytically shown to converge in the limit of pure-dephasing interactions between the central spin and the spin bath, with the couplings to the central spin being significantly larger than the intra-bath interactions $||\mathbf{A}||\gg||\mathbf{J}||$. The conditions appropriate for many systems of interest extend beyond the strict parameter ranges for which the CCE is guaranteed to converge, and several approaches have emerged in recent years that extend the CCE applicability to new, more complex systems and observables, some of which are described below.

First, the CCE method can be used to compute the autocorrelation of the Overhauser field upon the central spin~\cite{PhysRevB.90.115431, PhysRevB.92.161403}. In the secular approximation the autocorrelation function is given by:
\begin{equation}\label{eq:corr}
    \mathcal{C}_{AA}(t) = \left\langle\sum_{\{I\}} A_{zz}\hat I_z (t)\sum_{\{I\}} A_{zz}\hat I_z (0)\right\rangle,
\end{equation}
where the $\hat I_z (t)$ is the spin operator in the Heisenberg picture $\hat I_z (t)=\hat U^\dagger(t) \hat I_z \hat U(t)$. The autocorrelation function is factorized as:
\begin{equation}
    \mathcal{C}_{AA}(t) = \sum_{C} \Tilde{C}_{AA,\ C},
\end{equation}
where cluster contributions are defined recursively:
\begin{equation}
    \Tilde{C}_{AA,\ C} = {C}_{AA,\ C} -  \sum_{C'} \Tilde{C}_{AA,\ C' \subset C}.
\end{equation}

To improve the convergence of the cluster expansion for some challenging systems, one can directly sample the pure states of the bath to predict the decoherence in the thermal bath. Then, the coherence function can be computed as:
\begin{equation}
    \mathcal{L}(t) = \sum_i p_i\mathcal{L}_i(t),
\end{equation}
where $\mathcal{L}_i(t)$ is the coherence function computed for the pure bath state $i$, and $p_i$ is the probability of such state. In the high-temperature limit, all $p_i$ are equal. An even better convergence may be achieved by exactly averaging each cluster coherence function $L_C(t)$ over all possible pure states of a few nearby bath spins. Such an approach is sometimes referred to as hybrid CCE \cite{PhysRevB.86.035452,Yang_2016}.

For each pure bath state, one can compute the qubit states as eigenstates of the central spin Hamiltonian and include the mean-field interactions between the central and the bath spins; this approach enables the application of the CCE methods to study clock transitions of spin qubit systems \cite{PhysRevB.102.245303}.

An alternative way to improve the convergence of CCE calculations was presented in reference \cite{schatzle2024extended}, by redefining the set of clusters in the expansion to ensure the inclusion of strongly coupled pairs.

Another useful approach to understand spin coherence near avoided energy level crossings is the generalized CCE (gCCE) method, developed in Ref. \cite{yang2020gcce, PRXQuantum.2.010311}. Instead of projecting the total Hamiltonian on the qubit levels, one may directly include the central spin degrees of freedom into each cluster in the Hamiltonian. The coherence function of the cluster $L_C(t)$ is computed as $L_{C}(t) = \bra{0}\hat U_{C+S}(t)\hat \rho_{C+S} \hat U_{C+S}^{\dagger}(t)\ket{1}$,
where $\hat \rho_{C+S} = \hat \rho_{C} \otimes \hat \rho_S $ is the combined initial density matrix of the bath spins' cluster and central spin, and $\hat U_{C+S}(t)$ is the propagator that includes the full cluster Hamiltonian. 

The advantage of the generalized CCE is that it allows one to include the central spin Hamiltonian terms in a non-perturbative way, leading to increased accuracy in describing the dynamics of systems with a complex energy spectrum.

Finally, recent developments \cite{onizhuk2023understanding} have shown that the CCE is an approach sufficiently accurate even in the presence of a local dissipation on the bath spins, governed by the Lindblad master equation:
\begin{equation}\label{eq:lindbladian}
    \frac{d}{dt} \hat \rho (t) = \frac{-i}{\hbar} [\hat H, \hat \rho(t)] + \sum_i \gamma_i  \mathcal{D}[\hat L_i](\hat{\rho}),
\end{equation} 
where $\mathcal{D}[\hat L_i](\hat{\rho})\equiv
\hat L_i \hat \rho \hat L_i^\dagger - \frac{1}{2}\{\hat L_i^\dagger \hat L_i, \hat \rho \}$ are dissipations of the bath due to an external environment with short correlation time (much shorter than $1/\gamma_i$), and $\hat L_i$ are bath jump operators.

\begin{figure*}
    \centering
    \includegraphics[scale=1]{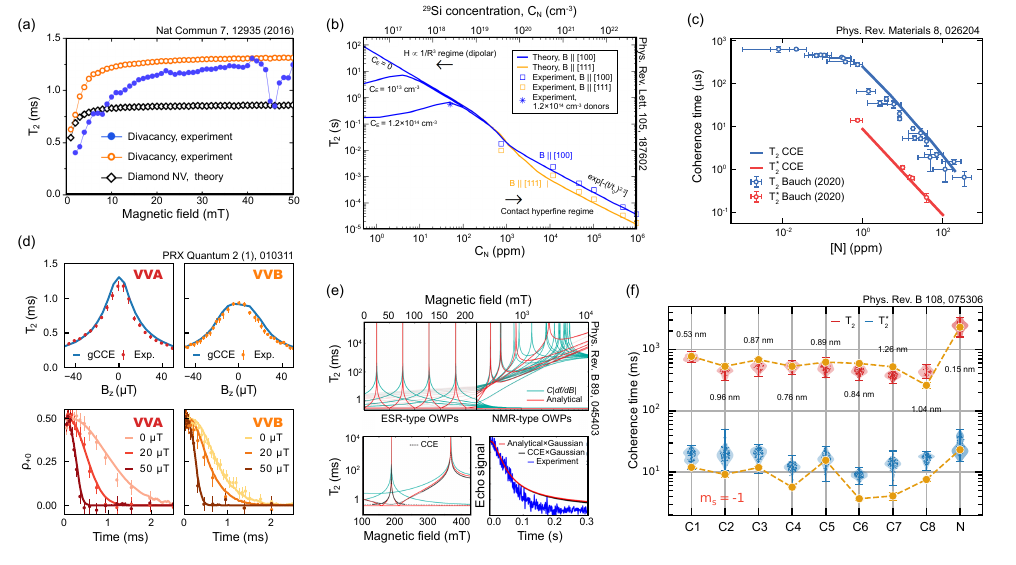}
    \caption{Validation of the results of cluster expansion methods with experimental data.
    (a) Experimental $T_2$ of the axial $kk$-$V_CV_{Si}$ divacancy spin ensemble in 4H-SiC as a function of magnetic field (B) (filled circles) compared with computed $T_2$ of the divacancy (empty circles) and computed $T_2$ of the \ch{NV^-} center in diamond (empty diamonds). Adapted from \cite{Seo2016} under CC 4.0 license. 
    (b) the values of $T_2$ of Si:P donor electron spins for various nuclear spin concentrations ($C_n$). At high $C_n$, contact hyperfine interactions are the dominant contributions and $T_2$ is dependent upon the magnetic field direction relative to the lattice orientation. At low $C_n$, $T_2$ is dependent upon the concentration of electron spins $C_E$. Experimental data are from \cite{PhysRevB.82.121201}. The figure is adapted from \cite{PhysRevLett.105.187602}. 
    (c) Computed values of $T_2$ and $T_2^*$ coherence times for various nitrogen $^{14}$N electron spin baths, overlaid with the corresponding experimental data from Ref. \cite{PhysRevB.102.134210}. The values of coherence times are extracted from a stretched exponential fit of the form $\exp[-(\sfrac{t}{T_2})^n]$. Adapted from \cite{marcks2023guiding}, copyright of American Physical Society. 
    (d) Single defect Hahn-echo coherence time of the basal $kh$-$V_CV_{Si}$ divacancy. Top: distribution of the values of $T_2$ for two different defects, VVA (left) and VVB (right), as a function of the magnetic fields $B_z$.
    Bottom: The Hahn-echo decay for three different values of the magnetic field. Solid lines correspond to theoretical predictions, and the points correspond to experimental measurements. Adapted from \cite{PRXQuantum.2.010311} under CC 4.0 license.
    (e) Top: $T_2$ values as a function of magnetic field for a variety of allowed transitions in Si:Bi, obtained using an analytical formula and gradient of the transition frequency. Bottom left: CCE predictions compared to an analytical formula. Bottom right: CCE predictions convolved with Gaussian $B$-field distribution of width 0.42 mT (arising from inhomogeneous broadening from the nuclear spin bath) provide an excellent fit to the measured Hahn echo decay around an ESR-type OWP \cite{Wolfowicz2013}. Adapted from \cite{PhysRevB.89.045403}.
    (f) Values of $T_2$ and $T_2^*$ of the nine nuclear spin registers measured by Bradley \textit{et al.}~\cite{PhysRevX.9.031045} in the vicinity of an \ch{NV^-} center in diamond, represented by yellow lines. Distributions correspond to CCE-computed coherence times $T_2$ (red) and $T_2^*$ (blue) when the NV is in the $m_s=-1$ state. Adapted from \cite{PhysRevB.108.075306}, copyright of American Physical Society.}
    \label{fig:validation}
\end{figure*}

\section{Spin decoherence in realistic systems with cluster methods}\label{sec:applications}

We now turn our attention to the setup of a realistic computational study using CCE to investigate the central spin coherence dynamics in the presence of spin baths.

After setting up the effective Hamiltonian (i.e. determining the terms to be included in the Hamiltonian, depending on the problem), one needs to determine the number of spins to be included in the CCE simulations. This number can be constrained by the geometry of the system (e.g. if one is interested in a single molecule the number of spinful atoms determines the number of spins), or it can be determined iteratively by progressively increasing the size of the bath until the convergence of results is achieved. For some spin defects in solids, the number of spins necessary for the convergence of the Hahn-echo signal has been reported to be of the order of several hundred (e.g. \cite{Seo2016}). This number can vary significantly, depending on the type of interactions, and reach more than tens of thousands of spins (e.g. \cite{PhysRevB.91.245416}).

Second, one needs to determine which clusters should be included in the expansion.
While the CCE method provides a way to reduce the scaling of the calculations with the number of bath spins from exponential to polynomial, in the majority of cases, it is not computationally feasible to include all possible clusters in the bath. Several procedures have been adopted to select the clusters to include (see, for example, \cite{PhysRevB.86.035452, Yang_2016}), and all of them imply the definition of a cut-off distance between bath spins determining whether two spins should be included in the same cluster. The value of such cut-off distance is determined iteratively, by checking the convergence of the desired results.

Third, the largest order in which  CCE calculations are carried out should be again determined iteratively, by verifying that the results are not significantly affected when adding clusters of larger size. In many dynamical simulations, second-order CCE calculations already provide accurate results (e.g. \cite{PhysRevB.85.115303, Bourassa2020}), but going to higher orders may be necessary for complex systems (\cite{PhysRevB.91.245416, Onizhuk2021}).

Finally, in the presence of numerical divergences or non-physical results, one should reevaluate the approximations chosen for the terms of the Hamiltonian and in the application of the CCE method itself.

The simple procedure outlined above may not be applicable to every problem but provides a simple framework that can be used to start tackling computational studies of a broad class of solid-state qubits, as we discuss below.

\subsection{Validation with experimental data}\label{sec:valid}

After appropriate verification of the numerical methods, any reliable computational study should address the validation of the results by experiments.

The CCE method has been widely applied to problems involving an electron spin in a nuclear spin bath. Nuclear spins constitute the dominant source of noise in many host materials, including GaAs~\cite{RevModPhys.79.1217}, Si \cite{Pla2012} and SiC \cite{Seo2016} with natural isotopic concentrations, and many other hosts. Such systems are perfect test beds for cluster expansion approaches, as they exhibit a strong backaction and thus cannot be described by assuming the presence of classical noise sources; in addition, they represent a coupling regime where the CCE approach is expected to converge (the coupling to the central electron spin is significantly larger than the coupling between the nuclear spins).

Consider the example of an electron spin in SiC. In Ref.~\cite{PhysRevB.90.241203} Yang \textit{et al.} predicted that electron spins in SiC have counter-intuitively larger coherence time than in the less spinful diamond host. Seo \textit{et al.} in \cite{Seo2016} further confirmed this prediction both experimentally and numerically with the CCE method, providing a robust validation of the CCE approach (Fig.~\ref{fig:validation} (a)). The qubits in isotopically purified 4H-SiC were further studied in Ref.~\cite{Bourassa2020}, where competing effects of nuclear and electronic spin baths were discussed, and in Ref.~\cite{PhysRevA.109.022603}, the authors provided a comprehensive study of the vacancy coherence in 3C-SiC. 

Another interesting example is that of phosphorous donors in Si. Shallow donors have a fairly delocalized wavefunction, and the contact term dominates the hyperfine interactions at long distances~\cite{PhysRevB.68.115322}. Witzel and Das Sarma~\cite{PhysRevB.74.035322} showed a good agreement between CCE simulations and the experimental measurements of the coherence time of the electron spin associated with phosphorous donors in naturally abundant silicon.

\begin{figure*}
    \centering
    \includegraphics[scale=1]{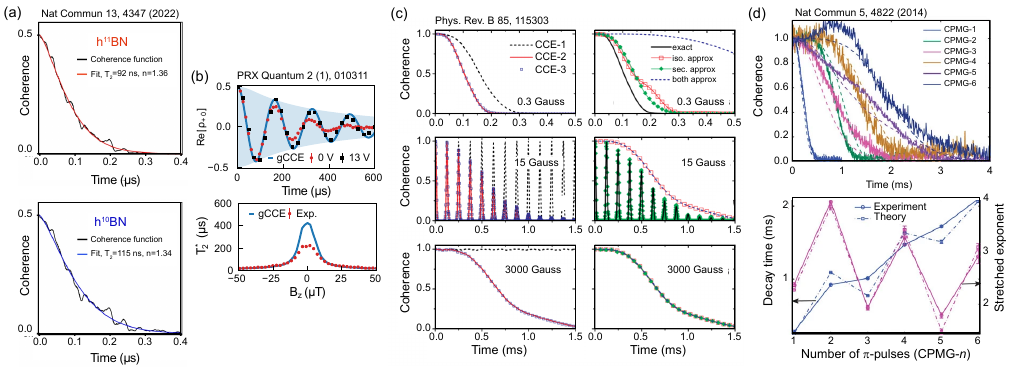}
    \caption{Interpretation of the experiments using cluster expansion methods. 
    (a) Simulated spin echo decay curve and corresponding stretched exponential fit for defects in h$^{11}$BN (top) and  h$^{10}$BN (bottom). Adapted from \cite{Haykal2022} under CC 4.0 license. (b) Top: Measured Ramsey signal of a single basal $kh$-$V_CV_{Si}$ divacancy at zero field with (black) or without (red) charge depletion, compared to theoretical predictions (blue). The shaded area corresponds to the theoretically predicted decay. Bottom:  $T_2^*$ as a function of the magnetic field ($B_z$). Adapted from~\cite{PRXQuantum.2.010311} under CC 4.0 license.
    (c) Decoherence mechanisms of Hahn echo signals of the \ch{NV^-} center in diamond under different magnetic field regimes. Left: The contributions to decoherence of different cluster sizes (CCE orders) under a magnetic field of 0.3, 15, and 3000 Gauss (top-to-bottom). Right: The contributions to decoherence of different types of interaction under a magnetic field of 0.3, 15, and 3000 Gauss (top-to-bottom). The red line with square symbols is the coherence calculated with the isotropic approximation of the hyperfine interaction. The green line with diamond symbols is the coherence calculated with the secular approximation. The blue dashed line is the coherence obtained with both the isotropic and the secular approximations. The black solid line indicates the exact results without approximations. Adapted from \cite{PhysRevB.85.115303} .
    (d) Top: Measured (solid lines) and calculated (dashed lines) coherence of the P-donor electron spin in the natural $^{29}$Si nuclear spin bath under CPMG control. The deviation seen at $\approx$ 1 ms for CPMG-6 is attributed to the overlap with uncorrected stimulated/unwanted echoes. Bottom: Comparisons of the experimental (solid lines) and theoretical (dashed line) decay times  (blue) and stretched exponents  (magenta) of the central spin decoherence under CPMG control. Adapted from \cite{Ma2014} under CC 4.0 license.
    }
    \label{fig:interpretation}
\end{figure*}
Further, Witzel \textit{et al.}~\cite{PhysRevLett.105.187602} also showed that the CCE could reproduce the experimental decoherence in the presence of paramagnetic impurities with varied concentrations, as long as one includes in the simulation the proper mean-field interactions with the spins outside the cluster~\cite{PhysRevB.86.035452} (Fig. \ref{fig:validation} (b)).

In a similar fashion, Park \textit{et al.}~\cite{Park2022} applied the CCE methodology up to the second order to compute the decoherence of the \ch{NV^-} center in diamond due to electron spins associated with nitrogen substitutional defects in diamond. Marcks \textit{et al.}~\cite{marcks2023guiding} further showed that converged CCE simulations accurately reproduce experimental data with varied concentrations of nitrogen in diamond~(Fig.~\ref{fig:validation}(c)).

In reference \cite{Du2009} the authors investigate the coherence of electronic spins in organic molecular crystals, specifically of spins formed in malonic acid under irradiation. The agreement between the predictions of cluster expansion methods and experimental results is excellent, under dynamical decoupling of the electron spin, and various decoherence channels present in the systems could be analyzed.

The interactions between the central electron spin and the spin bath can be greatly modified by fine-tuning the energy levels of the qubit with an external magnetic field. Specifically, in the presence of spin mixing terms in the Hamiltonian, such as a transverse ZFS $E$ or a strong hyperfine $A$ coupling (beyond GHz) to a single bath spin, the transition energy between two levels reaches a minimum at an avoided crossing, for a given value of the magnetic field. At avoided crossings, the frequency of the spin qubit does not depend on the magnetic field at first order, and the qubits thus are decoupled from the magnetic environment up to first order, leading to significantly improved coherence and performance \cite{PhysRevB.88.161412, Miao2020}. Spin transitions at avoided crossings are known as zero first-order-Zeeman (ZEFOZ) shifts~\cite{Miao2019}, clock transitions~\cite{Wolfowicz2013}, and optimal working points (OWP)~\cite{PhysRevA.90.042307}. The protected nature of the OWPs makes the spin qubits close to avoided crossing particularly challenging to simulate with cluster methods.

Balian \textit{et al.} \cite{PhysRevB.86.104428, PhysRevB.89.045403} showed that the CCE method can correctly capture the dynamics of the bismuth donor in silicon near OWPs. Bismuth has nuclear spin-$\sfrac{9}{2}$, that interacts strongly with the electron spin-$\sfrac{1}{2}$;  it exhibits a strong mixing of the spin levels at magnetic field strengths between 0-0.3 T. In this regime, multiple OWPs exist, as shown in Figure \ref{fig:validation}(e). Near the OWP, the Hahn-echo coherence time of the Bi donor in Si is increased by several orders of magnitude, and the results of CCE calculations convolved with qubit line-width broadening reproduce the experimental signals. We note, however, that there is a discrepancy between theory and experiment at longer timescales, indicative of the presence of some other noise source, not captured with the CCE calculations. 

Similarly, in electron systems with spin-1, clock transitions emerge at zero magnetic field if the transverse ZFS is sufficiently large. In this case, one needs to use the generalized CCE method to achieve a good agreement with experiments (Fig. \ref{fig:validation}(d)), as shown in Ref.~\cite{PRXQuantum.2.010311} for defects in SiC, where the comparison between theory and experiments is impressive.

We close this subsection with the example of the validation of CCE results in a significantly different scenario from those described above, where a nuclear spin which is near an electron spin qubit is considered as the central spin. In this case, a single electron spin in the bath dominates the dynamics of the environment.
Early studies of this problem in Silicon showed that the effect of more than $10^8$ spin pairs should be included in CCE calculations of order 2, to achieve convergence~\cite{PhysRevB.91.214303}.

Recent experiments have provided excellent reference results by mapping the positions, couplings, and coherence times of a large number of nuclear spins in the proximity of a defect in solids \cite{vandestolpe2023mapping, Abobeih2019, PhysRevX.9.031045}.
The CCE method can accurately describe the dynamics of single nuclei in such systems, e.g. matching the experimental values of the coherence time  $T_2$ of nuclear spins for different spin states of the \ch{NV^-} center in diamond (Fig. \ref{fig:validation}(f)) \cite{PhysRevB.108.075306}.

Overall, calculations with the CCE methods have provided results showing excellent agreement with experiments, even in a parameter space way beyond the conditions where the approximations underlying the methods are known to hold, giving us confidence in the CCE predictions for many complex systems.

\subsection{Interpretation of experiments}\label{sec:interp}

The problems that are commonly addressed by spin qubit coherence simulations fall broadly into two categories: (i) the characterization of the major sources of noise affecting known spin qubit platforms, and (ii) the prediction of properties of possible qubit systems not yet observed experimentally. In this subsection, we consider the first group of applications.

For example, in Ref.~\cite{Haykal2022}, the authors investigated the origin of the upper bound of a hundred nanoseconds to the coherence time of the Boron Vacancy in hBN at low magnetic fields; they established that it is mainly determined by the contact hyperfine interactions with the nearby nuclear spins (Fig.~\ref{fig:interpretation}(a)). In a similar fashion, in Ref.~\cite{chen2023long}, the authors identified the upper bounds for different sources contributing to the decoherence of spins in carbon nanotubes. The CCE method has also found applications in the interpretation of the hydrogen spin-induced decoherence of organic radicals in water \cite{Canarie2020, Jahn2022} and in biological systems \cite{Jeong2024}.

When investigating the dominant noise sources of spin qubits by including only the magnetic noise of the nuclear spin bath in the calculations, an unfavorable comparison with experiments would clearly reveal the presence of additional noise sources in the system. For example, the CCE calculations reported in Ref.~\cite{PRXQuantum.2.010311} showed a difference of a factor of two between computed and measured values of the $T_2^*$ of the basal divacancy in SiC at OWP; however, the calculations provided accurate predictions of the Hahn-echo $T_2$, clearly indicating the presence of an additional source of static noise. This source was identified as a charge noise that could be eventually negated by applying a constant voltage across the system (Fig. \ref{fig:interpretation} (b)).

Controlling the terms included in the general Hamiltonian allows one to determine which approximations to the spin dynamics hold in different magnetic field regimes (Fig. \ref{fig:interpretation} (c)).  Taking the \ch{NV^-} center in diamond as an example, Zhao, Ho, and Liu~\cite{PhysRevB.85.115303} showed that at weak magnetic fields, both secular and non-secular interactions of the central spin with a spin bath significantly contribute to the Hahn-echo signal; in addition, they showed that the dynamics of at least pairs of spin in the bath, corresponding to the CCE2 approximation, should be included in all regimes of the magnetic field. Finally, the authors provided a general framework for understanding the physical origin of the oscillatory behavior of the coherence function in different magnetic fields, providing a clear explanation of many experimentally observed phenomena. 

An interesting application of cluster expansion in characterizing noise is the study of a purely quantum spin bath as a noise source. For example, Zhao, Wang, and Liu characterized an anomalous decoherence effect in spin qubits with spin higher than one half \cite{PhysRevLett.106.217205}. By considering the spin levels with $\Delta m_s=2$ in the \ch{NV^-} centers as qubit states, they predicted coherence times longer than estimated with semiclassical models, due to the strong backaction of the qubit. This effect has been observed experimentally \cite{Huang2011}.

Another prediction then verified experimentally came from Kwiatkowski, Sza\'{n}kowski and Cywi\'{n}ski \cite{kwiatkowski2020}, who showed the emergence of a non-trivial phase in Hahn-echo experiment in a polarized spin bath, initially noticed by Paz-Silva, Norris, and Viola~\cite{PhysRevA.95.022121}. The predicted non-trivial Hahn-echo phase has been generalized to any type of quantum environment by Wang and Clerk~\cite{Wang2021}. These predictions were later confirmed experimentally by Jerger et al~\cite{PRXQuantum.4.040315}.

Using the LCE, Ma \textit{et al.} proposed a characterization technique that allows one to disentangle second-order and fourth-order correlations in a nuclear spin bath of the phosphorous donor in silicon (Fig. \ref{fig:interpretation} (d)) \cite{Ma2014}. Interpreting the dynamics of the bismuth donors in silicon, Balian \textit{et al.} investigated the interplay between the OWPs and dynamical decoupling protocols and found that in this case the dynamics of clusters of three bath spins must  be included into the simulation~\cite{PhysRevB.91.245416}, and Ma \textit{et al.} further showed that in this system the noise near an OWP  can be treated as an effective classical noise~\cite{PhysRevB.92.161403}.

The big challenge of both determining the validity of cluster approximations and their utility, lies in the interpretation of experimental data in the case of interface-dominated decoherence of solid-state spin qubit platforms. For example, in the case of the \ch{NV^-} center in diamond, there are numerous seemingly unconnected jigsaw pieces of experimental data \cite{PhysRevX.9.031052, PhysRevLett.122.076101}: some reports suggested an unexpectedly long correlation times of the static surface spins \cite{rezai2023probing}, others suggested spin hopping being the dominating contribution to the sensor spin coherence dynamics \cite{PRXQuantum.3.040328}.
Recent theoretical work \cite{candido2023interplay} also suggested that electric noise is a significant factor even at high magnetic fields.
The rise of experiments that can directly probe the local environment of surfaces with reporter spins \cite{PhysRevApplied.19.L031004} will provide valuable input for atomistic models of surfaces to be used in ab initio electronic structure calculations \cite{PhysRevApplied.20.014040, Zhu2023}. The latter, combined with accurate numerical simulations of coherence times, will, in turn, shed light on the effects dominating and ultimately determining the spin dynamics close to or at surfaces.

Finally, we mention another interesting avenue to explore with precise numerical simulations: the study of multiqubit systems \cite{PRXQuantum.2.030102}. Simulations using cluster methods showed interesting non-Gaussian effects in the nuclear spin-induced noise on a system of two-entangled NVs \cite{PhysRevB.98.155202}, as well as a robust entanglement between nuclear spin registers in spin defects \cite{maile2023performance}. We expect this area of research to grow in the upcoming years.

\begin{figure*}
    \centering
    \includegraphics[scale=1]{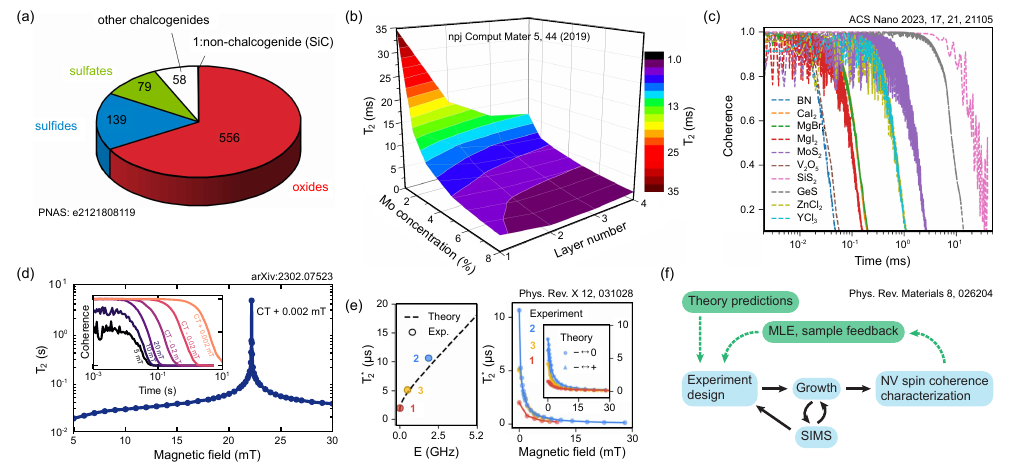}
    \caption{Predictions with cluster expansion methods. (a) Set of 832 stable compounds with predicted coherence time $T_2$ longer than 1 ms and predicted bandgap larger than 1.0 eV. SiC is the only stable widegap non-chalcogenide with $T_2>1$ ms. Adapted from \cite{Kanai2022} under CC 4.0 license.
    (b)  Isotopic engineering of two-dimensional materials. Computed coherence time $T_2$ (ms) of a few-layer MoS$_2$ as a function of Mo concentration. Here, both $^{95}$Mo and $^{97}$Mo have the same concentrations represented by the Mo concentration axis, and the concentration of $^{33}$S is 0.76\%, corresponding to its natural abundance. Adapted from \cite{Ye2019} under CC 4.0 license.
    (c) The calculated coherence for a representative defect within ten two-dimensional host materials. Adapted from \cite{Ali2023} .
    (d) Nuclear spin-limited $T_2$ of an electron spin near a clock transition (CT) of the \ch{Bi_{Ca}V_O^-} defect in CaO, as computed using the CCE method. The inset shows the actual computed coherence signal near a clock transition. Adapted from \cite{davidsson2024discovery} under CC 4.0 license.
    (e) Left: Zero-field spin coherence of three types of molecular qubits as a function of the transverse zero-field splitting along results obtained with the first-principles gCCE method (using the large $D$ limit). Right: Experimental and calculated $T_2$ as a function of the magnetic field. 1 corresponds to \ch{Cr(IV)(\textit{o}}-\ch{tolyl)_4} in \ch{Sn(\textit{o}}-\ch{tolyl)_4} matrix, 2 is \ch{Cr(IV)(\textit{o}}-\ch{tolyl)_4} in Sn(4-fluoro-2-methylphenyl)$_4$ matrix, 3 is Cr(IV)(2,3- dimethylphenyl)$_4$ diluted in Sn(2,3-dimethylphenyl)$_4$ matrix. Adapted from \cite{PhysRevX.12.031028} under CC 4.0 license.
    (f) Growth process workflow, incorporating CCE predictions. The traditional process steps (blue) for synthesizing a diamond NV sample. Iterations of growth and SIMS analysis are required to confirm nitrogen doping densities. The theoretical predictions and density maximum likelihood estimation (MLE) model (green) enable a non-destructive feedback process to circumvent SIMS and allow for an efficient experimental design. Adapted from \cite{marcks2023guiding}, copyright of American Physical Society.}
    \label{fig:predictions}
\end{figure*}

\subsection{Engineering new systems}\label{sec:pred}
As mentioned above, a second group of applications of quantitative numerical simulations of spin qubit decoherence involves predicting properties of interesting systems not yet available for experimental characterization.

Recently, Kanai \textit{et al.} provided a generalized scaling formula of the nuclear spin-limited Hahn-echo coherence time as a function of the nuclear spin species and concentration and electron spin type obtained with CCE simulations~\cite{Kanai2022}. The authors found that most host materials with expected high coherence time are oxides (Fig. \ref{fig:predictions}(a)). This work rejuvenated the interest in searching for promising spin defects in oxides, for example, by experimentally implanting well-known ions with a paramagnetic ground state into the new hosts \cite{zhang2023optical, dantec2021}, or by computationally carrying out searches using first-principles simulations~\cite{xiong2023}. In particular, the high-throughput search developed by Davidsson \textit{et al.}~\cite{Davidsson2021} allowed for the discovery of new defects in CaO with predicted coherence times beyond seconds at OWP(Fig. \ref{fig:predictions}(d))~\cite{davidsson2024discovery}.

Another rising field of theoretical predictions involves simulations of the properties of spin defects in 2D materials. Ye, Seo, and Galli predicted that isotopic engineering should be more efficient in 2D materials due to the reduced dimensionality of the host (Fig. \ref{fig:predictions}(b)) \cite{Ye2019}. Further, Lee \textit{et al.} proposed how to utilize isotopic purification and strain engineering in the case of the only 2D material with experimentally known spin qubits - hBN \cite{Lee2022}. Such isotopic engineering has been further shown experimentally to extend coherent control of the boron vacancy defect in hBN \cite{Gong2024}. There is, however, an inevitable effect of the substrate on the defect coherence times in 2D materials, as pointed out in Ref.~\cite{Onizhuk2021}. In this paper the authors also discussed the applicability of various theoretical approaches to recover the coherence dynamics in different geometrical confinements of the qubit.

In search of new van-der-Waals bonded host materials, in references \cite{PhysRevB.106.104108, Ali2023} the authors used the CCE in a high throughput search of defects in a large number of 2D hosts (Fig. \ref{fig:predictions}(c)). They provided a regression analysis of the spin coherence time as a function of the van-der-Waals host material properties to extract a scaling formula~\cite{PhysRevB.106.104108}, similar to the strategy adopted by Kanai \textit{et al.} in Ref.~\cite{Kanai2022}; in addition in Ref.~\cite{Ali2023}, the authors used a combination of DFT and CCE results to identify defect candidates for spin qubits in promising 2D materials.

One can further utilize cluster methods to quantitatively predict the properties of OWP in novel spin qubit systems. For example, the authors of \cite{PhysRevMaterials.5.074602} used the CCE method to predict the improvement of coherence time at clock transitions for the basal \ch{NV^-} centers in 4H-SiC, which is yet to be observed experimentally. 

Promising novel platforms for spin qubits are spins contained inside single molecules.
Optically addressable molecular qubits \cite{Bayliss2020} combine the high tunability of molecular compounds \cite{Laorenza2021} with the optical interface of the spin defects, making them promising candidates for quantum networking \cite{Laorenza2022}. By fine-tuning the structure of the molecule, one can significantly alter the qubit's coherence properties. In reference \cite{PhysRevX.12.031028}, the authors used the CCE method to predict the scaling of the Hahn-echo $T_2$ of molecular qubits as a function of the transverse ZFS at OWP; the predictions were corroborated with experimental data for three different compounds with slightly different compositions of the spin molecule and host crystal (Fig. \ref{fig:predictions}(e)). 

Numerical predictions can be further utilized as a characterization tool in materials' growth. In Ref.~\cite{marcks2023guiding}, the authors proposed the use of a library of coherence times generated with the CCE approach, together with a set of simple experimental pulse sequences, to recover the density of nitrogen substitutions in delta-doped diamond film samples (Fig. \ref{fig:predictions}(f)). Such an approach should be advantageous in scaling up the growth of spin qubit platforms.

As the field moves forward, we expect numerical simulations will be significantly more integrated into the experimental design of novel spin platforms, both in the search for new spin qubits as well as in addressing material science questions.

\section{Conclusions and Outlook}
As we hopefully showed above, the use of first-principles simulations to predict the decoherence dynamics of solid-state spin qubits opens opportunities for advancing both fundamental understanding and practical applications of solid-state spins in quantum technologies. As we continue to refine simulation methodologies, integrate new interdisciplinary approaches, and improve collaborations between theory, computation, and experiments, the field is poised to make significant strides toward realizing robust and scalable quantum technologies.

However, multiple issues remain to be addressed. For example, in the recovery of spin-phonon limited coherence, there is still no agreement between theoretical and experimental results at high temperatures for one of the best-studied systems, the \ch{NV^-} center in diamond (Fig. \ref{fig:phonons}(c)), indicating that higher-order perturbations play a significant role in the decay processes. Regarding spin-spin relaxation pathways, there remain major challenges in the characterization of the surface noise and of multiqubit systems, as well as in tackling spin baths regimes that require very high orders of the CCE method. Further perturbative treatments of the spin dynamics outside of a given cluster may help solve these problems, as well as potentially coupling the CCE with other methods to describe the dynamics of many-body quantum systems, such as tensor network approaches \cite{PhysRevLett.69.2863,  Montangero2018}.

In terms of emerging areas, we expect ab initio simulations to play a key role in interpreting the interface-limited decoherence of the near-surface solid-state spin qubits, as well as the behavior of multi-qubit systems. We further believe that the robust simulations highlighted in this work will be further used to optimize and predict the properties of future spin qubit platforms, including but not limited to novel spin defects, host materials, molecular qubits, and hybrid quantum platforms (e.g. Andreev levels-based spin qubits in superconducting devices~\cite{Hays2021}).
Finally, with the advent of open-source software for the simulation of spin dynamics~\cite{balian2015quantumbath, pycce}, we expect first-principles numerical predictions to become a valuable tool not only for computational scientists but also for experimental and theoretical groups aiming at validating measured data or analytical models.

\begin{acknowledgments}
This work benefited greatly from fruitful conversations with many scientists. In particular, we want to acknowledge Jonathan Marcks, Yu-Xin Wang, F. Joseph Heremans, and Tania Monteiro for useful comments on the manuscript. M.O. and G.G. acknowledge the support of NSF QuBBE Quantum Leap Challenge Institute (Grant No. NSF OMA-2121044) and the Design and Optimization of Synthesizable Materials with Targeted Quantum Characteristics (AFOSRFA9550-22-1-0370).
\end{acknowledgments}

\appendix
\section{A note on Bloch equations and quantum decoherence}\label{app:history}
Since the early magnetic resonance experiments in the second half of the twentieth century, the depolarization of the ensemble spins was classified into depolarization in the plane perpendicular to the external magnetic field (transverse magnetization) and parallel to the magnetic field (longitudinal magnetization). In the nuclear magnetic resonance (NMR) experiments, the dynamics of these processes is usually well-described by the classic Bloch equations \cite{PhysRev.70.460}:

\begin{align}\label{eq:bloch}
    \frac{dM_x(t)}{dt}&=\gamma (M_y(t)B_z(t) - M_z(t) B_y(t)) - \frac{M_x(t)}{T_2},\\
    \frac{dM_y(t)}{dt}&=\gamma (M_z(t)B_x(t) - M_x(t) B_z(t)) - \frac{M_y(t)}{T_2},\\
    \frac{dM_z(t)}{dt}&=\gamma (M_x(t)B_y(t) - M_y(t) B_x(t)) - \frac{M_z(t)-M_0}{T_1},
\end{align}
where $\mathbf{M}=(M_x,M_y,M_z)$ is the magnetization, $M_0$ is the thermal magnetization (assuming a constant magnetic field applied along the $z$-direction), $\gamma$ is the gyromagnetic ratio, $\mathbf{B}=(B_x,B_y,B_z)$ is the external magnetic field, and $T_1$ and $T_2$ are characteristic time constants. 
From Eq. (\ref{eq:bloch}), the expected depolarization behavior with time is represented by a pure exponential decay.
The transverse demagnetization with decay time $T_2$ is known as ``spin-spin relaxation" as it mostly stems from interactions with other spins; the longitudinal relaxation  $T_1$ is instead known as ``spin-lattice relaxation," and it is limited by the interaction of the spins with phonons in the crystal \cite{levitt2008spin}.

The same nomenclature was adopted in early papers exploring single solid-state spins for quantum information applications \cite{petta2005, PhysRevLett.92.076401, Fujisawa2002, PhysRevB.74.161203}, where the phenomenological relaxation times entering the Bloch equations correspond to the characteristic times of the {\it quantum} evolution of the spin-qubit. However, as discussed in the main text, the spin-qubit time dynamics and the qubit depolarization decay can be significantly more complex than just the exponential decay predicted with macroscopic Bloch equations.

\end{document}